\documentclass[journal,twoside,web]{IEEEtran}
\usepackage[compress,sort]{cite}
\usepackage{amsmath,amssymb,amsfonts}
\usepackage{algorithmic}
\usepackage{graphicx}
\usepackage{textcomp}
\usepackage{bm}
\usepackage{caption}
\usepackage{subcaption}
\usepackage[utf8]{inputenc} 
\usepackage[T1]{fontenc}    
\usepackage[hidelinks]{hyperref}     
\usepackage{url}            
\usepackage{booktabs}       
\usepackage{amsfonts}       
\usepackage{nicefrac}       
\usepackage{microtype}      

\hypersetup{
pdftitle={DCRA-Net: Attention-Enabled Reconstruction Model for Dynamic Fetal Cardiac MRI},
pdfsubject={eess.IV, cs.CV},
pdfauthor={Denis Prokopenko, David F.A. Lloyd, Amedeo Chiribiri, Daniel Rueckert, Joseph V. Hajnal},
pdfkeywords={Dynamic MRI Reconstruction, Fetal Cardiac MRI, Adult Cardiac MRI, Deep Learning, Attention Model},
}

\def\BibTeX{{\rm B\kern-.05em{\sc i\kern-.025em b}\kern-.08em
    T\kern-.1667em\lower.7ex\hbox{E}\kern-.125emX}}
\begin{document}

\title{DCRA-Net: Attention-Enabled Reconstruction Model for Dynamic Fetal Cardiac MRI}
\author{Denis Prokopenko, David F.A. Lloyd, Amedeo Chiribiri, Daniel Rueckert, and Joseph V. Hajnal
\thanks{
This work was supported by EPSRC Centre for Doctoral Training in Smart Medical Imaging (EP/S022104/1), Philips Medical Systems, Wellcome/EPSRC Centre for Medical Engineering WT (203148/Z/16/Z) and NIHR Biomedical Research Centre at Guy’s and St Thomas’ NHS Trust.}
\thanks{Denis Prokopenko is with King's College London, London, UK (e-mail: denis.prokopenko@kcl.ac.uk). }
\thanks{David F.A. Lloyd is with King's College London, London, UK and Evelina London Children’s Hospital, London, UK (e-mail: david.lloyd@kcl.ac.uk).}
\thanks{Amedeo Chiribiri is with King's College London, London, UK (e-mail: amedeo.chiribiri@kcl.ac.uk).}
\thanks{Daniel Rueckert is with Imperial College London, London, UK and Klinikum rechts der Isar, Technical University of Munich, Munich, Germany (e-mail: daniel.rueckert@tum.de).}
\thanks{Joseph V. Hajnal is with King's College London, London, UK (e-mail: jo.hajnal@kcl.ac.uk).}}

\maketitle

\begin{abstract}
Dynamic fetal heart magnetic resonance imaging (MRI) presents unique challenges due to the fast heart rate of the fetus compared to adult subjects and uncontrolled fetal motion. This requires high temporal and spatial resolutions over a large field of view, in order to encompass surrounding maternal anatomy.
In this work, we introduce Dynamic Cardiac Reconstruction Attention Network (DCRA-Net) - a novel deep learning model that employs attention mechanisms in spatial and temporal domains and temporal frequency representation of data to reconstruct the dynamics of the fetal heart from highly accelerated free-running (non-gated) MRI acquisitions.
DCRA-Net was trained on retrospectively undersampled complex-valued cardiac MRIs from 42 fetal subjects and separately from 153 adult subjects, and evaluated on data from 14 fetal and 39 adult subjects respectively.
Its performance was compared to L+S and k-GIN methods in both fetal and adult cases for an undersampling factor of 8x.
The proposed network performed better than the comparators for both fetal and adult data, for both regular lattice and centrally weighted random undersampling.
Aliased signals due to the undersampling were comprehensively resolved, and both the spatial details of the heart and its temporal dynamics were recovered with high fidelity.
The highest performance was achieved when using lattice undersampling, data consistency and temporal frequency representation, yielding PSNR of 38 for fetal and 35 for adult cases.
Our method is publicly available at \href{https://github.com/denproc/DCRA-Net}{github.com/denproc/DCRA-Net}.
\end{abstract}


\section{Introduction}\label{sec:introduction}

\IEEEPARstart{D}{ynamic} imaging of the fetal heart with MR presents a unique challenge.
The fetal heart rate is generally expected to be in the range of $110$-$170$ bpm~\cite{von2013normal} depending on the gestational age, compared to around $60$-$100$ bpm in adults.
This requires high temporal resolution to capture the rapid fetal cardiac cycle.
The small size of the fetal heart demands higher spatial resolution than for adults, to capture finer structures, while a large field of view is needed to encompass the maternal anatomy that surrounds the fetus.
In addition, any methods must be robust to incidental maternal and/or fetal motion.
These requirements result in a need for extremely rapid MRI k-space acquisition.
However, current acquisition rates of MRI scanners are too slow to sample the full k-space for large area of interest while maintaining high spatial and temporal resolutions.

To circumvent these limitations, a common approach in fetal dynamic MRI studies is to use reduced sampling of k-space during acquisition, with a complete imaging dataset generated via post-hoc reconstruction.
For example, this can be achieved by the continuous acquisition of the k-space signal, which is re-binned into cardiac phases using various gating techniques~\cite{feng2014golden,haris2017self,roy2013dynamic}, or by accelerated sampling followed by k-t SENSE reconstruction~\cite{tsao2003k} to form the initial estimation for further gating and outlier rejection~\cite{van2018fetal, van2019fetal, chaptinel2017fetal, roberts2020fetal}.

It is notable that fetal heart reconstruction methods are not as comprehensively studied as adult heart MRI reconstruction.
Previously, the application of convolution-based U-Nets~\cite{ronneberger2015u}, which utilised prior knowledge about data representation of dynamic MRI, highlighted the challenging nature of fetal heart~\cite{prokopenko2023deep, prokopenko2023challenge}.
The dynamic features of the fetal heart could be easily overlooked when using models with large receptive fields in the temporal and/or spatial domains, even when global metrics achieve competitive values.

\begin{figure*}[h!]
        \centering
        \includegraphics[width=\linewidth]{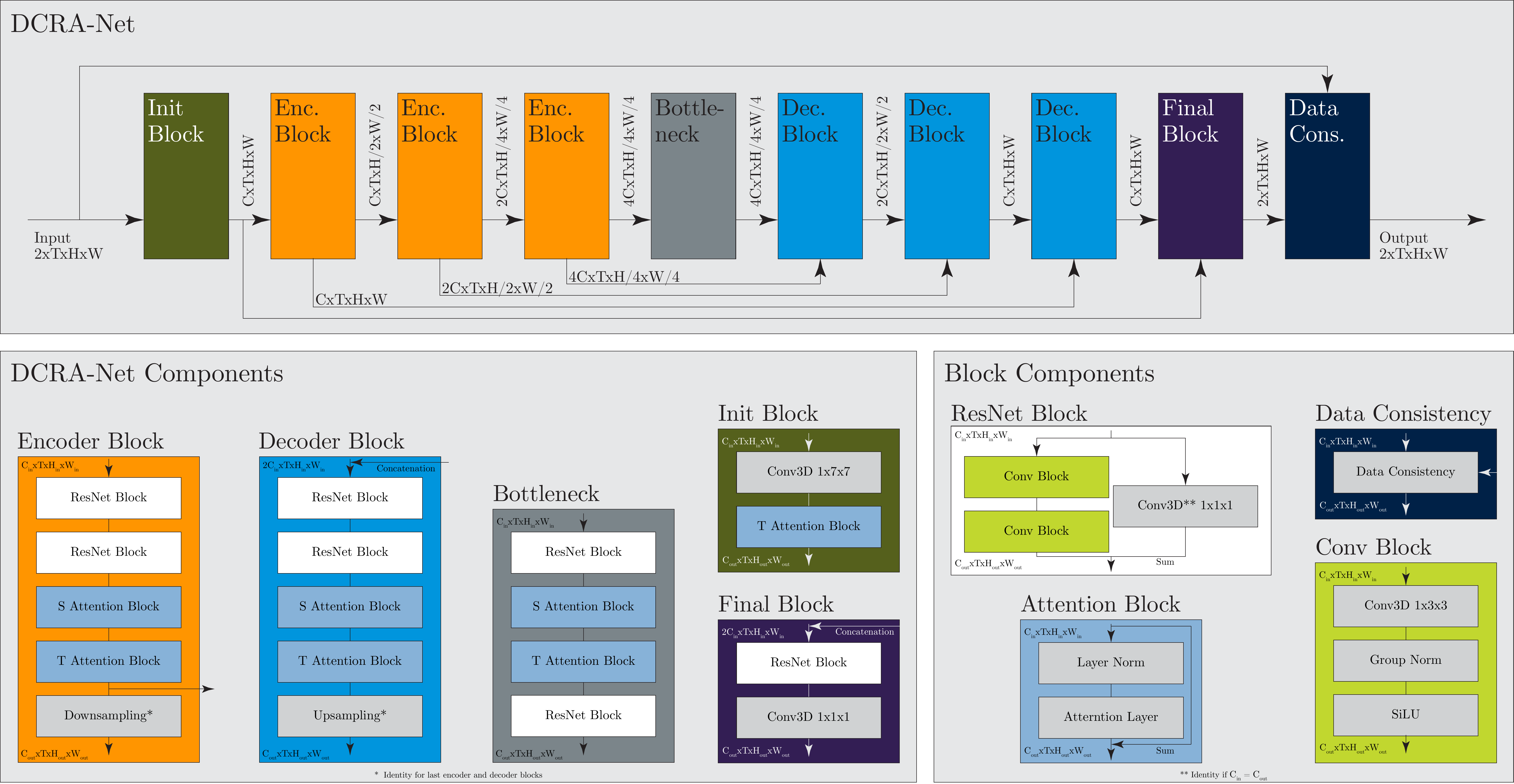}
        \caption{The proposed DCRA-Net model for dynamic fetal cardiac MRI reconstruction. The main parts of the model are encoder, bottleneck and decoder parts. Encoder and decoder blocks are built from two ResNet blocks, spatial (S) self-attention block followed by temporal (T) self-attention layer and down-/upsampling. The bottleneck consists of ResNet block, spatial and temporal self-attention layers followed by another ResNet block. The initial block uses spatial convolutional layer and temporal self-attention block. Our implementation assumes number of channels $C = 64$ for both fetal and adult application.}\label{fig:our_model}
\end{figure*}

In this work, we introduce Dynamic Cardiac Reconstruction Attention Network (DCRA-Net) - a deep learning reconstruction model that is able to recover the dynamics of fetal heart from highly accelerated free running (non-gated) MRI acquisitions, in an attempt to address the unique challenges of fetal heart imaging. 
We present a performance comparison with established methods in the context of our target domain of the fetal heart.
In addition, we evaluate the performance in application to adult cardiac MRI for a fairer comparison considering it is the native application domain of the comparator methods.
The key contributions of the current work are:
\begin{enumerate}
    \item We propose DCRA-Net, which is a 2D+time deep learning model to reconstruct accelerated non-gated dynamic fetal cardiac MRI.
    \item We evaluate the proposed DCRA-Net on our fetal and adult heart MRI data enabling fairer comparison with reference methods in application to both our target domain and their original data application.
    \item We provide an ablation study of DCRA-Net performance considering data representation, data consistency modes, acceleration rates ($4$ and $8$), and acquisition patterns (lattice and variable density random samplings).

\end{enumerate}

\section{Related Works}\label{sec:background}

\subsection{Conventional Reconstruction}

One of the first dynamic MRI reconstruction methods combined k-t space representation and Fourier encoding to recover unaliased image sequences at low acceleration rates ~\cite{madore1999unaliasing}, which was improved with the introduction of multi-coil processing~\cite{breuer2005dynamic,kellman2001adaptive}.
Access to spatiotemporal correlations and priors in k-t BLAST/SENSE~\cite{tsao2003k} improved quality and boosted acceleration rates.
The presence of substantially the same context throughout cardiac MRI examinations  allowed for low-rank or sparse representations, which were employed to enhance reconstruction in k-t FOCUSS~\cite{jung2009k}, k-t PCA~\cite{pedersen2009k}, k-t SLR~\cite{lingala2011accelerated}, L+S~\cite{otazo2015low}, and more advanced methods such as KLR~\cite{nakarmi2017kernel} and altGDmin-MRI~\cite{babu2023fast}.

\subsection{Deep Learning Based Reconstruction}

Recently, novel deep learning (DL) methods have unlocked new avenues for reconstruction methods optimised on large datasets. 
Deep Cascade of Convolutional Neural Networks (DC-CNN)~\cite{schlemper2017deep}, Convolutional Recurrent Neural Network (CRNN)~\cite{qin2018convolutional} and CINENet~\cite{kustner2020cinenet} were introduced as unrolled CNN models trained on retrospectively accelerated dynamic adult heart MR sequences using Cartesian sampling patterns.
For non-Cartesian radial and spiral sampling patterns, a variety of U-Net based models~\cite{jaubert2021real,jaubert2021deep, haji2021highly, wang2022deep} and unrolled networks~\cite{biswas2019dynamic, machado2023deep} were explored to suppress the artefacts caused by accelerated acquisitions.
Methods such as k-t NEXT~\cite{qin2019k}, LSNet~\cite{huang2021deep}, SLRNet~\cite{ke2021learned}, and CTFNet~\cite{qin2021complementary} combined properties of different representations of dynamic MRI and neural networks to enhance reconstructions following the iterative nature of the conventional reconstructions.
Some methods, like ME-CNN~\cite{seegoolam2019exploiting} and GRDRN~\cite{yang2022end}, employ the motion present in dynamic MRI as an additional feature to improve reconstruction and correct for movements.
Another advancement in dynamic MRI reconstruction methods came with the introduction of attention mechanisms~\cite{dosovitskiy2020image, vaswani2017attention}, which were integrated into transformer-based solutions such as RST~\cite{xu2023learning} and k-GIN~\cite{pan2023global}.

\section{Method}\label{sec:method}

Slice selective dynamic MRI can be expressed as a 3D volume with 2D k-space frames stacked over time, where each frame has two channels representing the real and imaginary parts of complex values.
This stacked, or time sequence, structure enables leveraging of the well-established 3D encoder-decoder architectures commonly used in medical applications~\cite{cciccek20163d,ronneberger2015u}, as well as more recent attention-based architectures designed for video generation tasks~\cite{dosovitskiy2020image, ho2020denoising, ho2022video,vaswani2017attention}.
Throughout this paper, we refer to the dimensions of the 3D volumes and their corresponding Fourier representations in the following order: temporal component, spatial height and width.

Figure~\ref{fig:our_model} shows the proposed DCRA-Net based on encoder-decoder architecture.
The model is factorised over spatial and temporal dimensions to decrease computational demands~\cite{arnab2021vivit}.
The initial and final blocks of the model transfer the input data representation into feature space and back.
The middle part of the model consists of encoder, bottleneck, and decoder.
Each encoder and decoder block consists of two ResNet blocks~\cite{he2016deep}, one spatial and one temporal self-attention layer~\cite{vaswani2017attention}, and a down/upsampling layer.
The ResNet blocks include two 3D convolutional layers with kernel size of $1 \times 3 \times 3$ to process only spatial dimensions.
The attention blocks process the data present in the corresponding spatial or temporal dimensions, while using the other as batch axis.
The downsampling and upsampling steps use convolutional and transposed convolutional layers with stride $2$ and kernel size $1 \times 4 \times 4$.
Finally, a data consistency layer~\cite{schlemper2017deep} is also included as an additional residual connection that passes input k-space data directly to the model output.

DCRA-Net can use input data with the temporal dimension represented in the time or frequency domains.
Following systematic testing (see ablation study in section~\ref{sec:results}) we will present most results for the temporal frequency representation of the input cardiac fetal and adult data.
A supporting rationale comes from the anatomical context, which remains static or slowly moves in most regions outside the heart, and so can be sparsely described in the temporal frequency domain, with signal concentrated in low temporal frequencies.

\section{Dataset and Experiments}

\subsection{Datasets}

\begin{figure}[t]
    \centering
     \begin{subfigure}[]{\linewidth}
         \centering
         \includegraphics[height=1in]{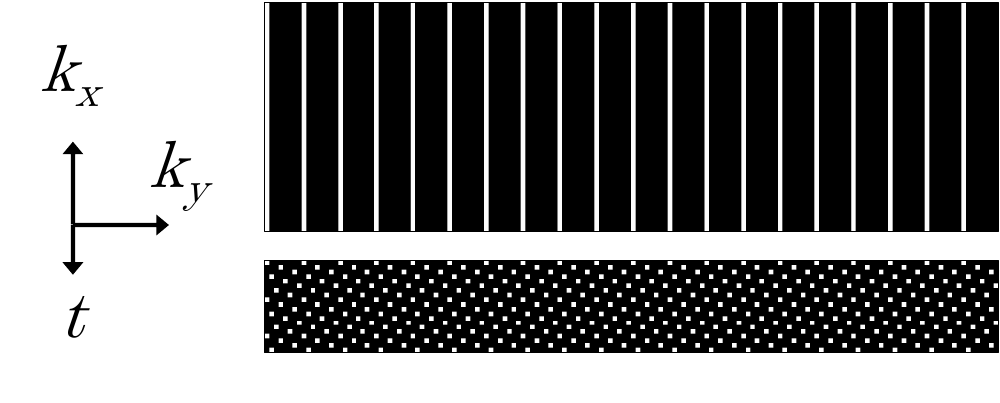}
         \caption{Lattice sampling pattern.} 
         \label{fig:lattice_sampling}
     \end{subfigure}
     \\
     \begin{subfigure}[]{\linewidth}
         \centering
        \includegraphics[height=1in]{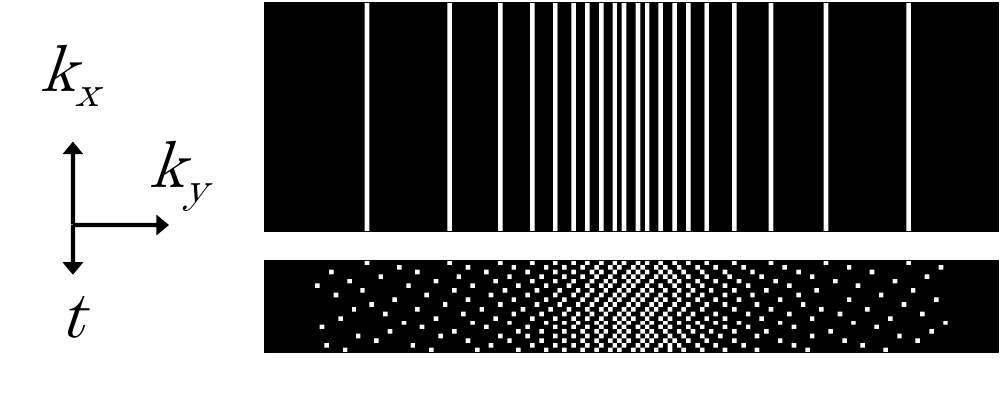}
        \caption{VISTA sampling pattern.} \label{fig:vista_sampling}
     \end{subfigure}
     \caption{Examples of $8$x undersampling patterns.} \label{fig:sampling_patterns}
\end{figure}

The fetal heart dataset consists of complex-valued images of non-gated dynamic MRI slices (6 mm thick) reconstructed from $8$x undersampled multi-coil k-space data using k-t SENSE~\cite{tsao2003k}.
The dataset contains $2267$ 2D slice + time volumes acquired from 56 subjects ($3$ healthy volunteers and $53$ patients with congenital heart diseases).
Informed consent was provided by each subject prior to their examination (ethical approvals 14/LO/1806, 07/H0707/105).
The dataset was collected using a Philips Ingenia $1.5$T MR system with balanced steady state free precession (bSSFP) sequence~\cite{carr1958steady} and $8$x accelerated regular Cartesian kt sampling pattern~\cite{tsao2005optimizing} on an underlying data grid of $152\times 400$ points.
Gestational age varies from $23$ to $35$ weeks.
The equivalent fully sampled spatial resolution  is $2.0 \times 2.0 \times 6.0$ mm, and the temporal resolution is $72$ ms per undersampled frame, which is sufficient for capturing fetal cardiac motion~\cite{roberts2020fetal}.
Each dynamic sequence contains $64$ or $96$ frames.

To enable fair comparisons with reference methods, which were proposed for adult heart MRI, and to explore how the proposed method functions in a generally familiar application domain, we have also made use of an adult heart dataset.
The adult dataset consists of complex-valued transverse images of gated, breath-held, dynamic cardiac MRI of the chest area of $192$ subjects (a total of 295 2D+time volumes).
The data was acquired on a Philips Achieva 3T MRI scanner using a QFLOW protocol, which provides single-coil single-slice cine reconstruction.
The ratio of male to female subjects is $93:99$ with ages spanning from $16$ to $82$ years (median age $51$).
The data was collected as part of clinically prescribed MRI sessions with subjects providing informed consent prospectively for their data to be used for research (approval 15/NS/0030).
The acquired voxel size is $1.25 \times 1.25 \times 8.00$ mm.
Each sequence was binned into $20$ cardiac phases of a single heartbeat.
The resolution of the reconstructed images varies from $174 \times 224$ to $311 \times 384$.

\begin{figure*}[t]
    \centering
    \includegraphics[width=\linewidth]{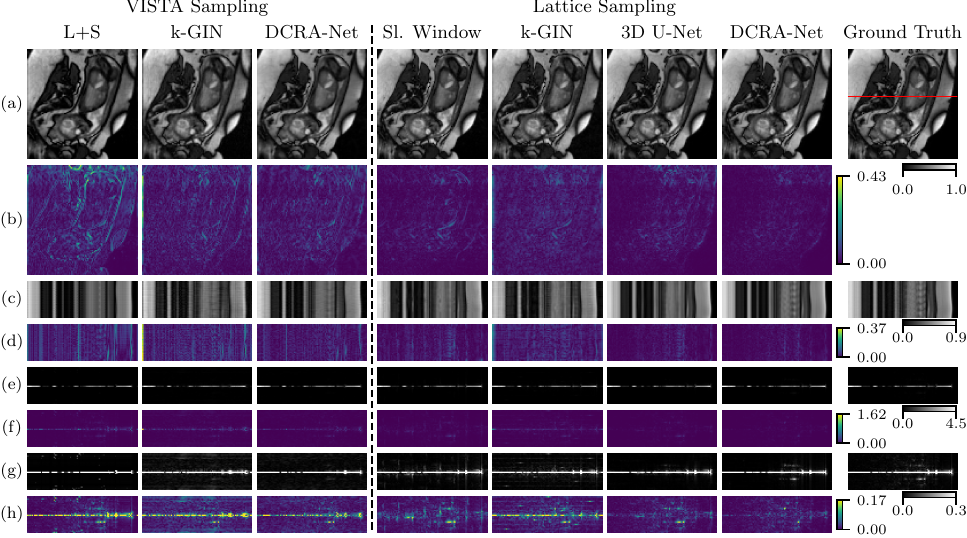}
    \caption{Fetal heart reconstruction comparison shows reconstructed data as image frames (a), temporal (c) and frequency (e) representations and their corresponding error maps (b, d, f). (e) and (f) are also shown using reduced dynamic range in (g, h).} \label{fig:sota_fetal}
\end{figure*}

\subsection{Implementation}

In this work, we use complex-valued image reconstructions as ground truth data for training the proposed model.
Complex values are represented as two-channel real-valued image pixel data.
The ground truth fetal data is scaled to a resolution of $96$ along the smallest spatial axis and centre-cropped to $96\times 96$ resolution for efficient use of computational resources.
To further optimize resource usage, the dataset was limited to $32$ frames.
Preliminary tests indicated that the choice of frame offset does not significantly impact reconstruction performance, so the first 32 frames were used.
The data was split into train-test subsets with $42/14$ patients corresponding to $1775/492$ slice sequences.
Adult data was resized to $160$ resolution along smallest axis and cropped to $160\times 160$ resolution.
The train-test splits contain $153/39$ patients resulting in $227/68$ slice sequences.

The proposed implementation of DCRA-Net encoder and decoder parts uses $3$ encoder and decoder blocks.
The initial feature representation has $C=64$ channels. 
The first encoder block maintains the same number of channels, while the subsequent blocks double the channel count.
On the decoder side, the first two blocks halve the number of channels, whereas the final decoder block retains the current channel count.
The self-attention layers have $8$ attention heads with feature size of $32$ per head.
The training procedure optimised mean absolute error between prediction and ground truth, with learning rate $10^{-4}$ over training cycle of $50$ epochs.

We train the proposed DCRA-Net on fetal and adult datasets independently.
The acceleration factor $8$ was chosen to match the acquisition strategy used for fetal data.
Retrospective undersampling of data was performed separately for lattice~\cite{tsao2005optimizing} and variable density incoherent spatiotemporal acquisition (VISTA)~\cite{ahmad2015variable} sampling patters shown in Figure~\ref{fig:sampling_patterns}.
We generated $100$ random VISTA sampling masks with the density parameter set to $0.7$, and Gaussian envelope set to $1/5$ of phase encoding dimension ~\cite{pan2023global, qin2021complementary}.
Other parameters of VISTA were left default.
For training, VISTA patterns were randomly subsampled from the generated set per sequence, while the same mask was used for testing models.

We evaluate model performance on $8$x accelerated fetal and adult data using normalised mean squared error (NMSE), peak signal-to-noise ratio (PSNR), and structural similarity (SSIM) from PyTorch Image Quality (PIQ) package ~\cite{kastryulin2019pytorch, kastryulin2022pytorch}.
The metrics were adapted to 3D complex-valued sequences and evaluated using the image representation of frames in the temporal domain.
Since these are global measures, we visually assess the heart dynamics for  fetuses and adults in each reconstruction.

In an ablation study, we systematically study the performance of the proposed DCRA-Net depending on representation (time or frequency) of the temporal component of the data used during training, and test the contribution of the data consistency to overall reconstruction quality.
This comparison is performed using fetal cardiac data using an $8$x undersampled lattice pattern.

We compare our DCRA-Net with conventional low-rank plus sparse matrix decomposition (L+S), 3D Convolutional U-Net and k-space Global Interpolation Network (k-GIN) reconstruction methods~\cite{otazo2015low, pan2023global, prokopenko2023challenge}.
The L+S was applied to $8$x accelerated data using VISTA masks.
The reconstruction parameters were $\lambda_L=0.277$, $\lambda_S=0.039$ and $\lambda_L=0.204$, $\lambda_S=0.057$ optimised on samples from fetal and adult training data respectively.
Convolutional 3D U-Net follows the implementation described in~\cite{prokopenko2023challenge} to reconstruct $8$x accelerated fetal data, which was undersampled using lattice pattern.
It uses $3\times 3\times 3$ convolutional kernels, $4$ downsampling steps, temporal frequency representation of the predicted data, data consistency and skip connection passing the average approximation of the input data. 
Implementation of k-GIN reconstruction follows description in~\cite{pan2023global}.
We trained k-GIN on both random and lattice sampling masks with acceleration factor $8$ on fetal and adult data from scratch for 300 epochs using the official code implementation.
To validate the results in the context of original k-GIN implementation, we also trained k-GIN and the proposed model on $4$x VISTA accelerated adult and fetal data and apply them to $8$x VISTA accelerated data. 

\section{Results}\label{sec:results}

\begin{table}[t]
    \centering
    \caption{Comparison with other methods for dynamic MRI reconstruction on $8$x accelerated fetal cardiac MRI showing mean and standard deviation values.}\label{tab:sota_fetal}
    \resizebox{\columnwidth}{!}{
    \begin{tabular}{| l |c  c c |} 
        \hline
        Model & NMSE $\downarrow$ & PSNR $\uparrow$ & SSIM $\uparrow$\\
        \hline
        \multicolumn{4}{|c|}{VISTA sampling}\\
        \hline
        L+S~\cite{otazo2015low} & $0.026 \pm 0.015$ & $26.882 \pm 1.988$ & $0.899 \pm 0.032$ \\
        
        k-GIN~\cite{pan2023global} & $0.017\pm 0.012$ & $29.509 \pm 3.837$ & $0.905 \pm 0.040$ \\
        \textbf{Proposed} & $\bm{0.010 \pm 0.006}$ & $\bm{31.065 \pm 2.211}$ & $\bm{0.934 \pm 0.020}$\\
        \hline
        \multicolumn{4}{|c|}{Lattice sampling}\\
        \hline
        Average & $0.016\pm 0.016$ & $30.086 \pm 3.698$ & $0.948 \pm 0.040$ \\
        Sl. Window & $0.007 \pm 0.011$ & $33.789 \pm 3.024$ & $0.951 \pm 0.030$ \\
        k-GIN~\cite{pan2023global} & $0.011 \pm 0.013$ & $31.896 \pm 4.448$ & $0.934 \pm 0.041$\\
        3D U-Net~\cite{prokopenko2023challenge} & $0.004 \pm 0.007$ & $36.628 \pm 3.273$ & $0.984 \pm 0.017$ \\
        
        \textbf{Proposed}  & $\bm{0.003 \pm 0.005}$ & $\bm{38.040 \pm 3.370}$ & $\bm{0.989 \pm 0.014}$ \\
        \hline
    \end{tabular}
    }
\end{table}

\begin{figure*}[t]
    \centering
    \includegraphics[width=\linewidth]{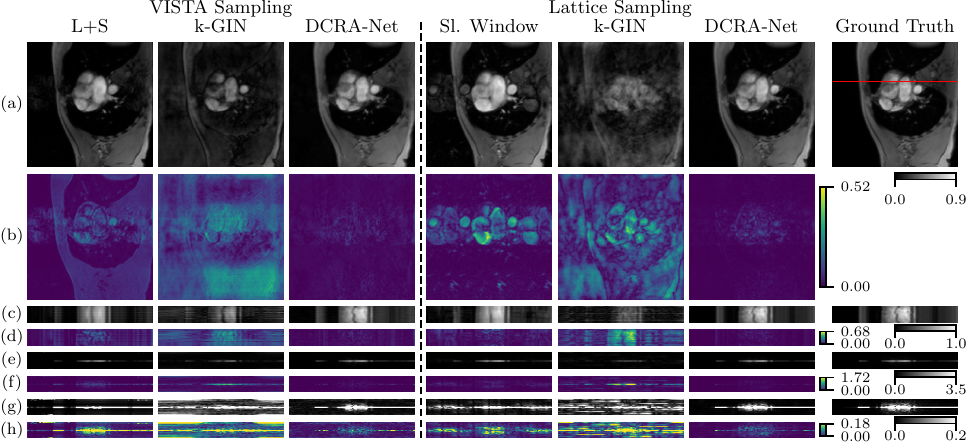}
    \caption{Adult heart reconstruction comparison shows reconstructed data as image frames (a), temporal (c) and frequency (e) representations and their corresponding error maps (b, d, f). (e) and (f) are also shown using reduced dynamic range in (g, h).} \label{fig:sota_adult}
\end{figure*}

DCRA-Net can process data represented either as time frames or temporal frequencies.  
We found in the ablation study (see below) that the latter resulted in superior performance, so will only present those results in detail.
The proposed DCRA-Net delivers the best reconstruction quality for fetal data accelerated with both VISTA and lattice undersampling patterns, as summarised for the evaluated metrics in Table~\ref{tab:sota_fetal} and visual results in Figure~\ref{fig:sota_fetal}.
Peak performance is achieved with lattice sampling pattern, yielding an SSIM value of $0.989$, outperforming all comparator methods.
For instance, the sliding window average, k-GIN and convolutional 3D U-Net deliver SSIM values of $0.951$, $0.934$ and $0.984$ respectively.
In Figures~\ref{fig:sota_fetal}c-d, the proposed model recovers temporal information about fetal heartbeat more closely aligned with the ground truth than the other methods.
Sliding window, k-GIN and convolutional 3D U-Net each capture the maternal breathing motion, but they fail to accurately reconstruct the fetal heartbeat.
The scaled temporal frequency representation in Figures~\ref{fig:sota_fetal}g-h evidences the presence of first frequency harmonics that characterise fetal cardiac motion, confirming the superior reconstruction quality of DCRA-Net.

For data undersampled using VISTA, the proposed DCRA-Net achieves similar performance, outperforming its competitors.
Evaluation results in Table~\ref{tab:sota_fetal} demonstrate that DCRA-Net surpasses both L+S and k-GIN.
The proposed model delivers SSIM values of $0.934$, which compares favourably to L+S and k-GIN SSIM values of $0.899$ and $0.905$.
Figures~\ref{fig:sota_fetal}c-h show that our model recovers motion information for both sampling strategies, with more precise reconstruction achieved using lattice undersampling.
In contrast, k-GIN recovers partial information about fetal heart motion only in the case of VISTA undersampling, as shown in Figure~\ref{fig:sota_fetal}c. 
However, the presence of other artefacts in the reconstruction affects the overall perception of the scan and reduces numerical scores compared to the lattice undersampling case.

\begin{table}[t]
    \centering
    \caption{Comparison of DCRA-Net with other methods for dynamic MRI reconstruction on $8$x accelerated adult cardiac MRI showing mean and standard deviation values.}\label{tab:sota_adult}
    \resizebox{\columnwidth}{!}{
    \begin{tabular}{| l | c c  c|} 
        \hline
        Model & 
        NMSE $\downarrow$ & PSNR $\uparrow$ & SSIM $\uparrow$\\
        \hline
        \multicolumn{4}{|c|}{VISTA sampling}\\
        \hline
        L+S & 
        $0.081 \pm 0.022$& $22.983 \pm 1.740$ & $0.763\pm0.040$\\
        
        k-GIN & $0.357\pm 0.057$ & $16.428 \pm 1.914$ & $0.431 \pm 0.036$ \\ 
        \textbf{DCRA-Net} & $\bm{0.009 \pm 0.005}$ & $\bm{32.523\pm 1.722}$ & $\bm{0.908 \pm 0.020}$ \\
        \hline
        \multicolumn{4}{|c|}{Lattice sampling}\\
        \hline
        Average & $0.043\pm 0.015$ & $25.820 \pm 1.423$ & $0.870 \pm 0.028$\\
        Sliding Window & $0.036 \pm 0.011$ & $26.935\pm 1.391$ & $0.857\pm 0.030$\\ 
        
        k-GIN & $0.562\pm0.088$ & $14.460\pm1.758$ & $0.277\pm 0.042$ \\
        \textbf{DCRA-Net} & $\bm{0.005\pm 0.003}$ & $\bm{35.047\pm 2.042}$ & $\bm{0.964\pm0.013}$\\
        \hline
    \end{tabular}
    }
\end{table}

As the reference methods were proposed for adult heart data, we analyse the reconstruction performance in their native domain for a more fair comparison (see Table~\ref{tab:sota_adult}).
The current DCRA-Net trained on $8$x accelerated adult data achieves SSIM values of $0.964$ and $0.908$ for lattice and VISTA sampling patterns respectively.
In contrast, the L+S and k-GIN deliver less reliable reconstructions resulting in worse values reported in Table~\ref{tab:sota_adult}.

The proposed DCRA-Net recovers most of the features of adult heart motion with minimal errors across the field of view, as illustrated in Figure \ref{fig:sota_adult}.
The temporal and frequency representations in Figures \ref{fig:sota_adult}c-h demonstrate the successful reconstruction of dynamic features of the heart, closely mirroring those present in the ground truth.
In contrast, L+S and k-GIN outputs on VISTA undersampled data introduce many artefacts across the field of view, preventing accurate depiction of heartbeat.
Moreover, the performance of k-GIN drops as we steer its application from VISTA to lattice undersampling pattern. 

\begin{table}[t]
    \centering
    \caption{Comparison of the proposed DCRA-Net with k-GIN showing mean and standard deviation values estimated on test dataset. The training is performed on $4$x VISTA accelerated data. The models were tested on $8$x VISTA accelerated data.}\label{tab:generalisation}
    \resizebox{\columnwidth}{!}{
    \begin{tabular}{| l | c  c c|} 
        \hline
        
        Model &
        NMSE $\downarrow$ & 
        PSNR $\uparrow$ & SSIM $\uparrow$\\
        \hline
        \multicolumn{4}{|c|}{$4$x Fetal Data}\\
        \hline
        k-GIN & 
        $0.008\pm 0.004 $ & $32.122 \pm 2.978$ & $0.934 \pm 0.026$ \\
        \textbf{DCRA-Net} & 
        $0.005\pm 0.002$ & $34.297 \pm 2.254$ & $0.964 \pm 0.012$\\
        \hline
        \multicolumn{4}{|c|}{$8$x Fetal Data}\\
        \hline
        k-GIN & 
        $0.017\pm 0.013$ & $29.487\pm 3.916$ & $0.906\pm 0.040$ \\
        \textbf{DCRA-Net} & 
        $0.180\pm 0.017$ & $17.975 \pm 1.976$ & $0.650 \pm 0.039$\\
        
        \hline
        \multicolumn{4}{|c|}{$4$x Adult Data}\\
        \hline
        k-GIN & $0.086 \pm 0.022$ & $22.842\pm 2.427$ & $0.726	\pm 0.044$\\

        \textbf{DCRA-Net} & 
        $0.002 \pm 0.003$ & $38.773 \pm 1.893$ & $0.959 \pm 0.013$ \\
        \hline
        \multicolumn{4}{|c|}{$8$x Adult Data}\\
        \hline
        k-GIN & $0.282	\pm 0.048 $ & $17.485\pm 1.924$ & $0.490\pm 0.038$ \\
        \textbf{DCRA-Net} & $0.056\pm 0.006$ & $24.449 \pm 1.339$ & $0.778 \pm 0.032$\\
        
        \hline
    \end{tabular}}
\end{table}

Table~\ref{tab:generalisation} shows evaluation of the proposed DCRA-Net and k-GIN models in the context of their generalisation abilities across $4$x and $8$x VISTA accelerations.
Both models deliver much better performance being trained and tested on $4$x accelerated data compared to results for training and testing using $8$x acceleration in application to both fetal and adult cases in Tables~\ref{tab:sota_fetal} and \ref{tab:sota_adult}.
The results on $8$x VISTA acceleration for k-GIN trained on $4$x match the reconstruction quality of the same architecture trained on $8$x undersampling.
The SSIM values on fetal data are $0.906$ and $0.905$ for k-GIN trained on $4$x and $8$x acceleration rates, while for adult case the results are $0.490$ and $0.431$.
In contrast, the proposed model has significantly reduced performance when deployed on acceleration rate unseen during training.
The proposed model trained on $4$x accelerated fetal data shows SSIM value of $0.650$ in application to $8$x accelerated data, which is worse than the SSIM of $0.934$ delivered by the same architecture trained on $8$x accelerated data.
A similar trend is observed in case of adult data as SSIM value drops from $0.908s$ to $0.778$.

\begin{figure}[t]
    \centering
    \includegraphics[width=\linewidth]{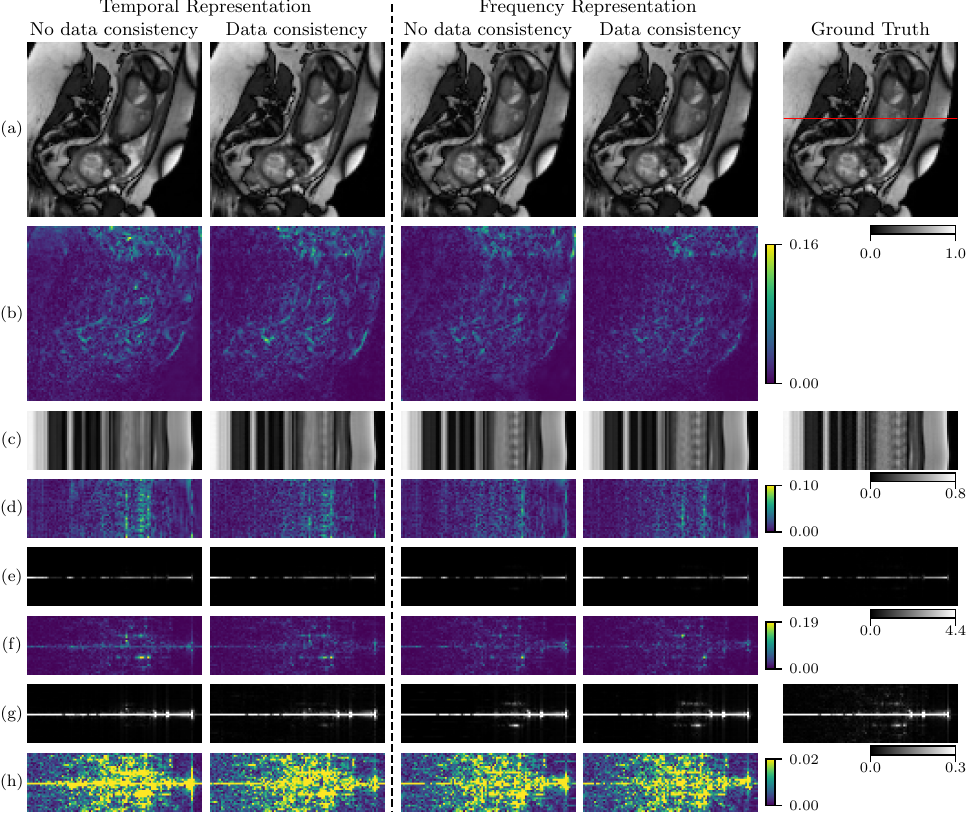}
    \caption{Comparison of DCRA-Net across temporal representations and data consistency presence. The figure shows reconstructed data as image frames (a), temporal (c) and frequency (e) representations and their corresponding error maps (b, d, f). (e) and (f) are also shown using reduced dynamic range in (g, h).} \label{fig:ablation_recon}
\end{figure}
\begin{table}[t]
    \centering
    \caption{The comparison of the proposed DCRA-Net in application to fetal data using different modes of data consistency (DC) and image data representation in temporal and temporal frequency domains.}\label{tab:ablation}
    \resizebox{\columnwidth}{!}{
    \begin{tabular}{| l | c  c c|} 
        \hline
        DC & NMSE $\downarrow$ & PSNR $\uparrow$ & SSIM $\uparrow$\\
        \hline
        \multicolumn{4}{|c|}{Temporal representation}\\
        \hline

        Disabled &  $0.004\pm0.007$ & $36.454\pm 2.980$ & $0.978\pm0.015$ \\
        
        Enabled  &  $0.003\pm0.006$ & $37.151\pm3.090$ &  $0.981\pm0.016$ \\
        \hline
        \multicolumn{4}{|c|}{Temporal frequency representation}\\
        \hline
        Disabled  &  $0.003\pm 0.006$ & $37.523\pm 3.238$ &  $0.988\pm0.014$ \\
        
        Enabled  &  $\bm{0.003\pm0.005}$ & $\bm{38.040\pm3.370	}$ &  $\bm{0.989\pm0.014}$ \\
        \hline
    \end{tabular}}
\end{table}

The ablation study of DCRA-Net applied to the fetal case demonstrates that both temporal frequency representation and data consistency (DC) contribute to improved reconstruction quality.
According to the numerical evaluation in Table~\ref{tab:ablation}, the model incorporating these features outperforms those using only temporal data representation, only DC or lacking both of them. 
However, the temporal frequency representation of data plays a more critical role, as it enables the reconstruction of fetal heart motion even in the absence of DC, as shown in Figure~\ref{fig:ablation_recon}.

\section{Discussion}\label{sec:discussion}

In this work, we address the challenge of fetal cardiac MRI reconstruction with Dynamic Cardiac Reconstruction Attention Network (DCRA-Net), which delivers the best reconstruction among tested options.
In the case of fetal heart imaging, cardiac motion is the most challenging feature to recover in the reconstructed scan.
Previously, the convolutional 3D U-Net~\cite{prokopenko2023challenge} did recover partial motion information, but this was far from the dynamics presented in the ground truth data.
DCRA-Net improves these results and reconstructs the dynamics of the heart more closely to the ground truth as shown in Figure~\ref{fig:sota_fetal}c.
The method recovers harmonics of the heartbeat, as revealed in the temporal frequency representation, which confirms the advance of this method.

The reference methods reconstruct only weak features of the motion or ignore them completely, while still delivering SSIM values around $0.9$.
This could be explained by noting that the fetal heart only occupies a small fraction of each frame and is moving rapidly, while there is a context of a full field of view that encompasses maternal anatomy.
Sliding window averaging recovers well structured image frames, showing SSIM value of $0.9515$, but leaves the fetal heartbeat completely unresolved.
This highlights not only the difficulty of fetal heart application, but also the need for appropriately nuanced evaluation that is sensitive to the target properties of application in addition to comparisons of global metrics.

The k-GIN shows less strong results than the proposed method for fetal cardiac MRI reconstruction as it does not fully recover the fetal heart motion.
However, it demonstrates generalisation across acceleration factors.
We observe that the k-GIN model trained on $4$x VISTA pattern and applied to $8$x VISTA undersampling performs similarly to the reported performance of k-GIN trained and tested on $8$x VISTA as shown in Tables~\ref{tab:sota_fetal}, \ref{tab:sota_adult}, and \ref{tab:generalisation}.
In contrast, the proposed DCRA-Net model recovers fetal heartbeat better for fixed acceleration, but its generalisation across different undersampling patterns is limited.

In this work we also used dynamic adult heart MRI as a reference application, since this is the target domain of many existing DL-based reconstruction methods~\cite{pan2023global,qin2021complementary,schlemper2017deep}.
The adult case shares some of the challenges of fetal cardiac MRI, but with a different balance of spatial occupancy of dynamic and more static image features, and slower temporal variation.
The results show that the proposed DCRA-Net, which was motivated by the challenges of fetal heart reconstruction, also performed well in an adult test case.
In Figure~\ref{fig:sota_adult}c, the proposed model recovers the heart cycle with minimal errors in both temporal and frequency representations (Figure~\ref{fig:sota_adult}d,h).  
Thus, focusing on fetal applications not only advances fetal heart imaging but can also hold promise for improving adult heart imaging and other dynamic MRI applications.
Unfortunately, in this study, L+S and k-GIN results were less successful than previously published performance~\cite{otazo2015low, pan2023global} in application to VISTA undersampling.
Perhaps the smaller size of our adult dataset compared to the fetal dataset and variation in sampled VISTA masks prevented us from fully harnessing the previously indicated potential of k-GIN to learn k-space structure.

Acquisition pattern plays a crucial role in dynamic MRI reconstruction as it defines sampled spatial frequencies of k-space.
The lattice pattern achieves uniform density across the whole of the desired k-space, capturing more higher spatial frequencies for a given undersampling factor than centrally weighted strategies such as VISTA.
Having access to such details seems to be important for the improved performance in application to the heartbeat reconstruction.
However, the uniform undersampling approach introduces more prominent, coherent aliasing artefacts that are more challenging for removal than noise-like artefacts.
Fortunately, both the proposed and reference methods succeeded in overall alias removal in the fetal case, but their performance in precisely reconstructing the fetal heart varies.

Sliding window averaging can achieve an approximation of fully-sampled k-space data for both types of undersampling studied here.
By definition, this  type of averaging method does not recover all the temporal details, but it can serve as a powerful baseline (alias free) approximation, which is closer to the ground-truth than the outputs from L+S and k-GIN methods in Figure~\ref{fig:sota_fetal}.
This highlights the importance of the visual assessment of the results, as the values of the evaluated metrics suggest great reconstruction quality, although the dynamic content is clearly underrepresented.

The proposed DCRA-Net demonstrates strong performance in both fetal and adult heart reconstruction, but its application is currently limited to single-coil data, similar to the k-GIN~\cite{pan2023global} approach.
Demonstration on such single-coil data is helpful in exploring concepts in deep learning models for fetal (and general) cardiac MRI reconstruction, paving the way for more complete multi-coil data approaches.
Likewise, in common with many past studies, the present work relied on target unaliased reconstructions, rather than native undersampled data for training and testing. 
Within these constraints assumption, we managed to provide crucial insights on relative performance at a lower cost, while maintaining feasibility of the experiments from computational perspective.
Application to raw multi-coil data needed to achieve a full clinical imaging pipeline will be the subject of future work.

\section{Conclusion}\label{sec:conclusion}

In this work, we introduced DCRA-Net - a model for dynamic fetal heart MRI reconstruction that utilises an encoder-decoder structure with factorised attention mechanisms, temporal frequency representation and data consistency.
The model successfully reconstructed the dynamics of the fetal heart from heavily accelerated data, demonstrating its potential to achieve higher spatial and temporal resolutions, which is crucial for addressing the challenges posed by fast heartbeats and small heart structures in fetal subjects.
The proposed model outperformed L+S, k-GIN, and convolutional 3D U-Net, achieving superior numerical global metric values.
Visual assessment confirmed that all tested solutions successfully captured general fetal and maternal movements, but only DCRA-Net demonstrated significant improvement in the reconstruction of the fetal heart, more accurately recovering the periodic heart motion.
Although motivated by the challenges of fetal cardiac MRI, our experiments demonstrated that the proposed model outperformed the reference models when applied to adult cardiac MRI data, even though this was the original target domain for the comparator methods.
Such adaptability suggests that the proposed method has potential for broader applications in dynamic MRI reconstruction applications.
Future work will aim to expand the proposed DCRA-Net to multi-coil data containing complementary information across the coils, which is needed for application to prospectively acquired raw data.

\bibliographystyle{ieeetr}
\bibliography{main.bib}

\end{document}